\documentclass[english,conference]{IEEEtran}
\usepackage[T1]{fontenc}
\usepackage[latin9]{inputenc}
\usepackage{amsthm}
\usepackage{amsmath}

\providecommand{\tabularnewline}{\\}

\theoremstyle{plain}
\newenvironment{lyxcode}
{\par\begin{list}{}{
\setlength{\rightmargin}{\leftmargin}
\setlength{\listparindent}{0pt}
\raggedright
\setlength{\itemsep}{0pt}
\setlength{\parsep}{0pt}
\normalfont\ttfamily}%
 \item[]}
{\end{list}}

\usepackage{hyperref}

\usepackage{babel}

\begin{document}

\title{The Boost.Build System}

\author{\IEEEauthorblockN{Vladimir Prus}\IEEEauthorblockA{Computer Systems Laboratory\\ Moscow State University, CS department\\ Moscow, Russia\\ vladimir.prus@gmail.com}}
\maketitle
\begin{abstract}
Boost.Build is a new build system with unique approach to portability.
This paper discusses the underlying requirements, the key design decisions,
and the lessons learned during several years of development. We also
review other contemporary build systems, and why they fail to meet
the same requirements.
\end{abstract}

\section{Introduction}

For software projects using compiled languages (primarily C and C++),
build system is the key element of infrastructure. Mature tools such
as GNU Make\cite{make} or GNU Automake\cite{autobook} exist. However,
those tools are relatively low level, and hard to master. They also
have limited portability. For that reason, many software project do
custom work on build system level. One such project is C++ Boost \cite{boost}.

C++ Boost is a popular collection of C++ libraries, many of which
are either included in the next revision of the C++ Standard or planned
for inclusion\cite{c++std,tr1}. This unique position attracts a lot
of users, who, in turn, use a wide variety of operating systems and
differently-configured environments. This differs from most commercial
projects --- which target a few platforms that are important from
business perspective, and are built in a well-controlled environment.
This is also different from most open-source projects --- which tend
to focus on GNU/Linux environment. Developers' background also considerably
differs --- a person who is expert in C++ is not necessary an expert
in different operating systems. 

This diversity in user and developer base lead to the following requirements
for a build system:\label{requirements}
\begin{enumerate}
\item {}``Write once, build everywhere''. If a library builds on one platform,
it should be very likely that it builds on all other platforms supported
by the build system. It follows that build description should be relatively
high-level, and avoid any system- or compiler- specific details, such
as file extensions, compiler options or command-line shell syntax
details.
\item Extensibility. Adding support for a new compiler or a platform should
not require changing build system core or build descriptions for individual
libraries. Ideally, user would have to write a new module and provide
it to the build system.
\item Multiple variants. The build system may not require that the build
properties for the entire project are specified up-front, during special
{}``configure'' step, and then require that all build products be
removed before changing build properties. Instead, it should be possible
to change build properties without full rebuild. Such change may happen
at three levels:

\begin{enumerate}
\item Between different parts of the project. The simplest example is compiling
a specific source file with an additional compiler option. More complex
example is building a specific module as a static library, and another
as a shared library. It should be possible to change every aspect
of the build process --- even including the used compiler.
\item Between different builds. For example, one may originally build a
project in release mode for testing and, after discovering a bug,
wish to initiate a build in debug mode. It would be wasteful to remove
the previously built object files, so the build system must arrange
for debug and release products to be placed in different directories.
The mechanism should not be restricted to just debug and release builds,
but apply to any build properties. 
\item Within one build. This means that one invocation of the build system
may produce several variants of a build product --- for example, static
and shared versions of the same library.
\end{enumerate}
\end{enumerate}
This paper describes Boost.Build\cite{boostbuild} --- a build system
developed to meet the above requirements. Section 2 will describe
key concepts and mechanisms of Boost.Build. In section 3 we review
the lessons learned during development as well as some unexpected
drawbacks. Section 4 discusses other contemporary build tools, and
why they could not be used. Section 5 summarizes the article and suggests
future development directions.

\section{Design concepts}

The best way to explain the key design elements of Boost.Build is
by following a few steps of gradual refinements, starting from a classic
tool --- GNU Make. In GNU Make, a user directly defines a set of \emph{targets},
where a target is an object that has:
\begin{itemize}
\item a name
\item a set of \emph{dependency targets}
\item a command to build the target
\end{itemize}
Consider this example:
\begin{lyxcode}
a.o:~a.c
\begin{lyxcode}
g++~-o~a.o~-g~a.c
\end{lyxcode}
\end{lyxcode}
Here, the name of the target is \texttt{a.o}, the only dependency
is a target named \texttt{a.c}, and the command invokes the \texttt{gcc}
compiler. Given the set of targets defined in a build description
file ({}``buildfile'' for short), GNU Make identifies targets that
are out of date with respect to their dependencies, and invokes the
commands specified for such targets. The description shown above has
two problems. First, the names of the targets and the exact commands
typically may vary depending on environment, and should not be hardcoded.
This problem is typically solved using variables --- in the example
below, the OBJEXT and CFLAGS variables may be defined as appropriate
for platform.
\begin{lyxcode}
a.\$(OBJEXT):~a.c
\begin{lyxcode}
g++~-o~a.o~\$(CFLAGS)~a.c
\end{lyxcode}
\end{lyxcode}
While this makes build description more flexible, it also makes it
rather verbose, and hard to write. Second, depending on build variant
and platform, even the set of targets may vary. For example, depending
on platform and desired linking method, building a library might produce
from 1 to 4 files. Obviously, conditionally defining 4 targets for
every library is extremely cumbersome.

Modern build systems do not require that user describes concrete targets,
but provide a set of \emph{generator functions}, or \emph{generators}
for short. A generator is called with the name of primary target,
a list of sources, and other parameters, and produces a set of concrete
targets. Sometimes, these concrete targets are GNU Make targets. Sometimes,
a different low-level build engine is used. For example, a library
might be defined like this%
\footnote{For presentation purposes, we have abstracted away the syntax of modern
build systems.%
}:
\begin{lyxcode}
library(helper,~helper.c)~~
\end{lyxcode}
This statement calls a function \texttt{library} that constructs the
set of concrete targets that is suitable for the target platform ---
which may include the library file proper, import library, and various
manifest files. The compiler and linker options are also derived from
both the platform and the way the build system was configured. Thus,
the user-written build description does not include any platform-specific
details. Instead, such details are handled by the build system, which
is separately maintained. Boost.Build also uses a similar description
mechanism, but advances it further.

First key observation is that using generators is not sufficient to
achieve portability. Requirements listed in section 1 include using
different build properties for different parts of the project. This
can be achieved using additional parameters to generators. But if
the set of those parameters, or their values, is either incomplete,
or depends on platform, the build description is not portable. To
achieve the portability goals, Boost.Build defines a large set of
build parameters with the following characteristics:
\begin{itemize}
\item Every generator accepts the same set of build parameters
\item The names of build parameters and their values are the same everywhere
\end{itemize}
For example, every generator in Boost.Build accepts a parameter named
\texttt{optimization}, with \texttt{none}, \texttt{speed} and \texttt{space}
as possible values. Consequently, the example might be modified as
follows:
\begin{lyxcode}
library(helper,~helper.c,~

~~~~~~~~~~~~optimization=space)~~
\end{lyxcode}
This change addresses requirement 1 ({}``write once, build everywhere'').

The second key observation is that differences between platforms are
so significant that creating a single generic generator such as \texttt{library}
is hard. Obviously, small behaviour differences can be handled in
an ad-hoc way --- for example by introducing a global variable set
by platform-specific code and checked by the generator. However, in
existing tools there are dozens of such variables, with the generator
still containing significant platform specific logic. More systematic
approach is needed. To that end, Boost.Build allows several generators
to exist, and uses a dispatching function to select the generator
to use. In example below:
\begin{lyxcode}
library(helper,~helper.c,

~~~~~~~~~~~~link=shared)
\end{lyxcode}
the description written by the user looks the same as before. However,
the \texttt{library} function is no longer responsible for constructing
targets. Instead, it merely selects and invokes a platform-specific
generator. This generator need only deal with a single platform, and
can be easily implemented. The specific generator selection algorithm
(that will be described below) allows new generators to be defined
in platform-specific modules and automatically participate in generator
selection, thereby addressing requirement 2 ({}``extensibility'').
It should be noted that recursive calls are common --- for example,
the \texttt{library} generator might use the \texttt{object} generator.
In Boost.Build, the dispatching function is also used for such recursive
calls, allowing for fine-grained customization. 

The third key observation is that if a build description is allowed
to call a dispatching function when the build description is parsed,
it severely limits the possibilities to further build the same part
of a project with different properties. To address this issue Boost.Build
introduces \emph{metatargets} --- which are essentially closure objects.
Consider an example using the actual Boost.Build syntax:
\begin{lyxcode}
lib~helper~:~helper.cpp~;
\end{lyxcode}
This statement defines a closure of the dispatching function, binding
the name and the sources list. If we invoke Boost.Build from the command
line using the following command:
\begin{lyxcode}
\$~b2~toolset=gcc~variant=debug

link=shared
\end{lyxcode}
then the closure object will be called with the specified build parameters.
The \texttt{toolset=gcc} and \texttt{link=shared} parameters uniquely
specify a generator --- \texttt{gcc.link.dll} --- that is called to
produce the concrete targets. In the example below, we request a two-variant
build:
\begin{lyxcode}
\$~b2~toolset=gcc~link=shared~-{}-

toolset=msvc
\end{lyxcode}
In this case, the created closure object will be called twice, once
with \texttt{toolset=gcc }and\texttt{ link=shared} parameters, and
once with \texttt{toolset=msvc} parameter. Different generators will
be selected, and a substantially different set of concrete targets
will be produced. The metatargets mechanism addresses requirement
3 ({}``multivariant builds'').

We have introduced the key design elements of Boost.Build. The remainder
of this section describes in detail the most important mechanisms
used to implement this design.

\subsection{Requirements}

It is uncommon for the entire project to be buildable for all possible
build parameters. \emph{Requirements} is a mechanism to restrict the
possible build parameters for a specific metatarget. Simple requirement
merely state that a given build parameter should always have a specific
value for this metatarget. For example:
\begin{lyxcode}
lib~helper~:~helper.cpp

~~~~:~link=static~;
\end{lyxcode}
overrides the value of the \texttt{link} build parameter that was
passed to the metatarget, and causes the concrete targets to be constructed
as if \texttt{link=static} was passed. Conditional requirements override
a build parameter if some other parameters have specific values. For
example:
\begin{lyxcode}
lib~helper~:~helper.cpp~

~~~~:~toolset=msvc:link=static~;
\end{lyxcode}
will override the \texttt{link} build parameter only if the \texttt{toolset}
build parameter has the value of \texttt{msvc}. Finally, indirect
conditional requirements specify that a user-provided function should
be called to adjust build properties. 

For convenience, a buildfile may specify \emph{project requirements}
that are automatically added to requirements of all metatargets in
that buildfile.

\subsection{Platform support}

This section explains two mechanisms that facilitate easy support
for new platforms --- selection of generators by the dispatching function,
and translation of build parameters into properties of concrete targets.

Let's look again at the syntax used to declare a metatarget:
\begin{lyxcode}
lib~helper~:~helper.cpp~;
\end{lyxcode}
As said before, this creates a closure of the dispatching function,
binding target name, list of sources, and --- which we did not say
before --- the metatarget type, in this case \texttt{lib}. For the
purpose of generator selection, Boost.Build maintains additional information
about each generator --- the metatarget type, and the set of required
build parameters. For a concrete example, consider the following table:

\begin{tabular}{|c|c|c|}
\hline 
Generator & Type & Required parameters\tabularnewline
\hline
\hline 
gcc.link.dll & LIB & toolset=gcc\tabularnewline
\hline 
gcc.link & EXE & toolset=gcc\tabularnewline
\hline 
msvc.link.dll & LIB & toolset=msvc\tabularnewline
\hline
\end{tabular}

When the dispatching function is called, it first selects the generators
associated with the metatarget type. In our example, such generators
are \texttt{gcc.link.dll} and \texttt{msvc.link.dll}. Then, required
parameters of the selected generators are compared with the build
parameters passed to the dispatching function. If any of the required
parameters is not present, the generator is not considered. In our
example, if \texttt{toolset=gcc} is passed to the dispatching function,
then \texttt{msvc.link.dll} generator is discarded. All the remaining
generators are called. If exactly one succeeds in generating targets,
then the dispatching function returns. Otherwise, an ambiguity is
reported and the build process stops. This selection mechanism allows
additional generators to be easily added, without modifying core logic
of the build system.

When a specific generator constructs a target, it should establish
the exact path and name of the target, as well as the command to build
it. All that typically depends on build parameters. Of course, a generator
may use arbitrary logic to compute this information, but Boost.Build
comes with convenient default behaviour. The target path is constructed
using the values of build parameters. For example, a path might be
\texttt{bin/gcc/debug/}. Some mechanisms are used to make the paths
shorter --- for example, for a few common parameter the path includes
only the values, but not the names. Also, parameters that have default
values are not included in path. Target name is constructed from the
name of the corresponding metatarget. Boost.Build maintains a table,
that is indexed by metatarget type and the value of the \texttt{target-os}
build parameter, and gives suffix and prefix that should be added
to metatarget name. The mechanism to construct updating command is
the key to easy definition of new generators, and is illustrated below:
\begin{lyxcode}
actions~gcc.link.dll~\{

~~~~g++~-shared~\$(OPTIONS)

\}

flags~gcc.link.dll~OPTIONS~

~~~~:~<profiling>on~:~-pg~;
\end{lyxcode}
First, a \emph{command template} is defined --- by convention, it
has the same name as a generator. Command template may refer to variables,
in this case \texttt{OPTIONS}. The second statement in the example
establishes mapping between build parameters and variables that are
replaced in command template. Given these declarations, a generator
can create a new target specifying \texttt{gcc.link.dll} as command
template for that target. All the \texttt{flags} statements for this
command template are automatically processed. The \texttt{flag} statement
above requests that if build parameter \texttt{profiling} has value
\texttt{on}, then the \texttt{-pg} string be added to the \texttt{OPTIONS}
variable. After all flag statements are processed, every reference
to a variable in the command template is replaced by the variable's
value. This mechanism proved to be highly beneficial, because it allows
to add support for a new build variable to any generator with a very
localized change.

\section{Current state and learned lessons}

At this point Boost.Build is a mature tool that can be successfully
used in production environment, and has already met its requirements.
At the same time, it is being actively developed. This section will
describe the main issues that were discovered.

\subsection{Metatarget-induced indirection}

In most existing build tools, buildfiles are written in some interpreted
language, and are executed at build system startup, calling generators
and constructing targets. Boost.Build differs from this model by creating
closure objects that are called with proper build parameters at a
later point. Furthermore, Boost.Build does not require that project
be {}``configured'', with some of the build parameters fixed, before
starting a build. Consequently, when a buildfile is executed it does
not make sense to talk about {}``current'' build parameters and
no logic that depends on build parameters may be implemented as an
\texttt{if}-statement on the top level of a buildfile. Rather, such
logic must be implemented as functions that will be called by generators.

For many users, this trait cause understanding problems. We believe
that this complexity directly follows from the requirements and key
design elements, and cannot be fully eliminated. On the other hand,
most users successfully adjusted to this model.

\subsection{Code-level extension mechanisms}

One of the goals of Boost.Build was simple description language. This
lead to invention of concise syntax for many tasks. However, often
no suitable programmatic interface was designed for the case when
the concise syntax is not enough. In other words, there are many areas
where build behaviour needs to be customized by the user, and there's
a wide spectrum of possible customization mechanisms --- from a new
build parameter to a new metatarget type. In a few cases, this spectrum
is not evenly covered, and user has to choose between a very simple
method that is not flexible enough, and an extremely complex solution.

One example is the conditional requirements syntax shown previously:
\texttt{toolset=msvc:link=static}. This syntax is sufficient for the
majority of cases, but does not support complex conditions --- in
fact, conditions using any logical operators except for {}``and''.
Until indirect conditional requirements were introduced relatively
late during development, users were forced to use a fairly verbose
mechanism instead.

Another example is generators. It is very easy for user to declare
a new generator that produces one output target from one input target.
However, any conditional logic --- such as creating an additional
target depending on some build parameters --- requires substantial
complexity. While we have described generators as functions in this
article, they are actually implemented as classes in certain programming
language, which adds some overhead for just declaring a new generator.
Furthermore, the implementation of the base generator classes was
not designed for easy extensibility, so often, user had to reimplement
significant amount of code.

We believe that issues of this kind have only small correlation with
the key design choices, and can be eliminated. In fact, quite a few
were already fixed as user report them.

\subsection{User expectations}

One unexpected issue during development was users' expectations. It
is safe to say that most users either have GNU Automake background,
or are not experienced with command line tools, and these users have
some specific, and often different, expectations.

For example, GNU Automake allows to change compiler by setting environment
variables, such as \texttt{GXX}. Users often try the same with Boost.Build,
and find that it has no effect. For another example, Boost.Build does
not stop after a compile error, but builds other targets that do no
depend on the failed one. At the end of the build, a summary of failures
is printed. This small change proved problematic. Many users did not
understand that the error was printed earlier, and interpreted the
summary as the original error. And on some operating systems, finding
an error in several thousand lines of build output is a problem itself.
Developers on Microsoft Windows operating system usually expect that
every tool checks system registry for all configuration. Consequently,
they found it very unnatural when prior versions of Boost.Build required
to specify compiler location in a configuration file. Finally, Boost.Build
command line syntax is slightly unusual, having separate syntax for
command line options, build parameters and arguments. Many users still
try to use option's syntax to specify build parameters.

Some of those issues are natural consequences of a different design
and require users to adjust. But still, many issues are independent,
and can be easily addressed. We recommend that design process for
any project in an established area include explicit gathering of user
expectation to avoid unnecessary differences in operation details.

\section{Existing solutions}

There are two build systems that are most commonly used today -- the
one integrated with Microsoft Visual Studio, and the Automake build
system. However, neither of them is truly portable. Below, we review
a few solutions that work across different platforms.

\subsection{Eclipse CDT}

The C/C++ Development Tooling (CDT) for the Eclipse Platform comes
with its own build system\cite{managedbuild}. The CDT build system
keeps a repository of available tools, organized in named toolsets.
For each tool, input and output file types are specified. Any project
is required to specify the name and type of the target that should
ultimately be built, and CDT automatically picks tools that can produce
the desired final target from all files in the source folders. Each
tool can have a set of options that are editable via user interface.
Similar to Boost.Build, tool options may be specified on individual
source folders or individual source files.

Let's review how CDT build system can support the build system requirements
described in section\ref{requirements}:
\begin{itemize}
\item The {}``write once, build everywhere'' requirements is not met.
User can count on some functionality to be available everywhere ---
in particular shared and static libraries and predefined {}``debug''
and {}``release'' build variants. However, any further fine-tuning
is done via options that are specific to each tool. Therefore, if
building with a different compiler, the options has to be specified
anew. CDT actually has an indirection level between the value of an
option as displayed in user interface, and the command line flags
used for compilation when that value is selected. Therefore, it would
have been possible to implement a portable set of options, that every
tools would translate into appropriate command line options. However,
since such portable set is not defined, build descriptions in CDT
are not portable.
\item The extensibility requirement is poorly met. The only way to extend
build system is via new tool definition, and it not possible to completely
override the build process for a specific platform. Further, tool
definition should be be included either in CDT core or in a separate
Eclipse plugin, and cannot be easily packaged with a project. Tool
definition uses an XML-based language, which appears to be inconvenient
in practice.
\item The multivariant requirement is partially met. CDT supports build
configurations that include a complete set of options for all tools,
and allows one to freely change build configuration every time a build
is initialized. Each part of a complex project may use different build
configuration. Note that this happens because a {}``project'' in
CDT terminology may only contain a single final target, so non-trivial
user project has to be split into multiple CDT projects. It is not
possible to build the same target in different configurations simultaneously.
It is also not possible to build using arbitrary ad-hoc set of build
parameters --- one has to define a new configuration instead.
\end{itemize}
CDT has one unique feature --- it has a mechanism to specify dependencies
between build parameters. For example, 64-bit compilation can be enabled
only if the chosen processor indeed supports 64-bit instructions.
This feature can be worthwhile to implement in Boost.Build.

\subsection{CMake}

The CMake \cite{cmake,cmake_site} build tool is designed as a abstraction
on top of existing {}``native'' build systems. When invoked, CMake
reads its buildfiles, and then generators secondary buildfiles using
a selected {}``backend'' build system -- for example, GNU make.
The project is then build when the user explicitly runs the secondary
build system. Whenever CMake buildfiles are modified, or any build
properties must change, the secondary buildfiles are regenerated.
Such scheme improves build system speed for the case where a project
is repeatedly build with the same settings.

Let's review how CMake can support the build system requirements described
in section\ref{requirements}:
\begin{itemize}
\item The {}``write once, build everywhere'' requirement is not met. The
points we have raised when discussing CDT equally apply to CMake.
\item The extensibility requirement is not met. Platform-specific modules
in CMake essentially specify variables used by CMake core, and it
is not possible to completely replace generators%
\footnote{The {}``generator'' is used in the sense defined in this article.
CMake documentation uses the word {}``generator'' for an unrelated
concept.%
}. Support for some platforms, for example, Microsoft Visual Studio,
relies on special core functionality.
\item The multivariant requirement is not met. CMake requires that a project
is configured with specific set of properties, and requires reconfiguration
for any change in properties. This appears to be result of an explicit
design goal, meant to make it impossible to accidentally mix modules
built with incompatible settings. 
\end{itemize}

\subsection{SCons}

The SCons build tool\cite{scons,scons_article} is unique in two aspects.

First, it uses the Python language for buildfiles, as well as implementation
language. In contrast, both Boost.Build and CMake use custom languages.
On one hand, this means that the syntax is not as concise, due to
punctuation and quoting rules of Python. On the other hand, the use
of a mature and widely known programming language reduces learning
curve, and simplifies many programming tasks. It also means there's
no language boundary between buildfiles and build system core.

Second, SCons uses cryptographic signatures to detect if a target
should be rebuild. When a target is built, signature of content of
all source files, as well as the command used to produce the file
is stored. These signatures are recomputed on every build, and if
they differ from the scored ones, the target is rebuild. This approach
means that a change in command associated with a target will be detected,
and cause a rebuild, while other build system can build different
parts of project with incompatible commands. 

For the simple usage, SCons provides a set of generators that can
be called from buildfiles, for example \texttt{Program} and \texttt{Library}.
The targets produced by those generators use globally specified build
parameters. However, SCons also provides a mechanism called \emph{construction
environment} --- explicitly created collection of build parameters.
It is possible to invoke a generator in specific environment, consequently
building different parts of projects with completely unrelated set
of build properties. However, this mechanism does not allow to easily
build one target with different properties. SCons does not assign
different directories for target with different properties, so user
is required to explicitly specify different names for final and intermediate
targets.

Let's review how SCons supports the requirements described in section\ref{requirements}:
\begin{itemize}
\item The {}``Write once, build everywhere'' requirement is not met. The
points we have raised when discussing CDT equally apply to SCons.
\item The extensibility requirements is met. A platform specific code can
completely replace standard generators if so desired.
\item The build variants requirements is partially met. Similar to CDT,
SCons allow to explicitly define several build variants, and unlike
CDT, all variants can be built simultaneously. However, SCons does
not automatically place products for different parameters in different
directories, so it is in general not possible to change any parameter
between two build invocation without discarding previous build products.
\end{itemize}

\section{Conclusions and Future Work}

This paper has presented the requirements for the Boost.Build system
and its key design decision, as well as reviewed some existing solutions.
We believe that the main distinguishing characteristics of Boost.Build
are:
\begin{itemize}
\item portable build properties, and associated mechanisms like requirements
\item true multivariant builds, specifically the metatarget concept
\item convenient extensibility, in the form of generator selection and flags
mechanisms
\end{itemize}

While some design complexities were encountered, we believe that overall,
Boost.Build is a step forward in the area of software construction.

There are two key areas of future development:
\begin{itemize}
\item Use of the Python language for implementation and build description.
We find that Python has become sufficiently popular and well-supported
and the benefits of using it will outweight slightly more verbose
syntax.
\newpage
\item IDE integration. Because Boost.Build does not rely on any legacy backend
build tools, and because every metatarget can be repeatedly constructed
with different, or same, build properties, it is particularly suitable
for integrated development environments --- making it possible to
quickly determine what products must be rebuild as result of changes
made by a user.
\end{itemize}

\section{Acknowledgments}

The author would like to thank Alexander Okhotin and Sohail Somani
for reviewing drafts of this paper.

\bibliographystyle{IEEEtran}
\bibliography{IEEEabrv,references}

\end{document}